\documentclass[twocolumn,showpacs,amsmath,amssymb,prl]{revtex4}

\usepackage{graphicx}% Include figure files
\usepackage{bm}% bold math
\begin{document}

\title{Resonance energy and charge pumping through quantum SINIS contacts}
\author{N. B. Kopnin$^{(1,2)}$, A. S. Mel'nikov$^{(3)}$, and V. M. Vinokur$^{(4)}$}
\affiliation{$^{(1)}$ Low Temperature Laboratory, Helsinki
University of
Technology, P.O. Box 2200, FIN-02015 HUT, Finland,\\
$^{(2)}$ L. D. Landau Institute for Theoretical Physics, 117940
Moscow, Russia\\
$^{(3)}$ Institute for Physics of Microstructures,  603950, Nizhny
Novgorod, GSP-105, Russia\\
$^{(4)}$ Argonne National Laboratory, Argonne, Illinois 60439}

\date{\today}

\begin{abstract}
We propose a mechanism of quantum pumping mediated by the spectral
flow in a voltage-biased SINIS quantum junction and realized via
the sequential closing of the minigaps in the energy spectrum in
resonance with the Josephson frequency. We show that the dc
current exhibits giant peaks at rational voltages.
\end{abstract}

\maketitle

\begin{figure}[t]
\centerline{\includegraphics[width=0.7\linewidth]{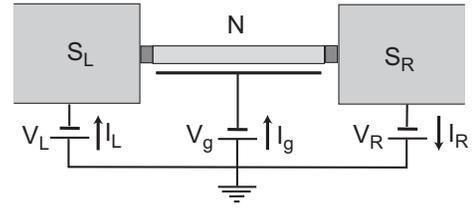}} %
%\begin{figure}[t] %h=here,t=top,b=bottom,p=erilliselle kuvasivulle
%\centerline{\epsfxsize= 6cm\epsfbox{trajectories.eps}}
\caption{Charge pump through a quantum channel with a variable
chemical potential. The bias voltage is $V=V_L-V_R$.}
\label{fig-quantdot}
\end{figure}

Quantum pumping in mesoscopic structures offers a unique
possibility for studying and direct manipulating the fundamental
quantum characteristics of the nanoscale objects (see, for
example, \cite{AltshulGlazm,Switkes,Buttiker} and references
therein).  Most of the previously proposed quantum pumps
(including the superconductor--quantum-dot junction of Ref.\
\cite{Hekking}) used the cyclic adiabatic processes in order to
avoid the relaxation which was believed to smear out the quantum
behavior~\cite{Buttiker}.  In this Letter we propose and
investigate a realization of the {\it nonequilibrium resonance
charge pump} in a form of a superconductor-normal-superconductor
(SNS) junction which essentially relies on both the discrete level
dynamics due to the adiabatic variation of scattering parameters
of the SN contacts and on the relaxation processes in the
continuum quasiparticle spectrum.  We discuss an exemplary device,
the symmetric voltage-biased SINIS structure (see Fig.\
\ref{fig-quantdot}) consisting of two superconducting leads (S)
coupled via the (tunnel) barriers (I) and the quantum ballistic
normal conductor (N). The dc current (the pumped charge) exhibits
giant peaks at rational bias voltages provided the chemical
potential of the normal conductor is varied in certain compliance
with the Josephson frequency. This requires a nonequilibrium
electron distribution achieved under the condition that the
Josephson frequency is higher than the inelastic relaxation rate
but smaller than the Andreev levels spacing in the normal
conductor. The relaxation of this distribution is of critical
importance since it enhances the pumping efficiency to the degree
that the dc current peaks greatly exceed the rectification current
(Shapiro steps) observed in the equilibrium state of the same
junction, thus providing unambiguous manifestation of the quantum
pumping effect (see the discussion in Ref.~\cite{Buttiker}).

%%%%%%%%%%%%%%%%%%%%%%%%%%%%%%%%%%%%%%%%%%%%%%%%%%%%%%%%%%%%%%%%%%%%%
%%%%%%%%%%%%%%%%%%%%%%%%%%%%%%%%%%%%%%%%%%%%%%%%%%%%%%%%%%%%%%%%%%%%%%

\begin{figure}[t]
\centerline{\includegraphics[width=1.0\linewidth]
{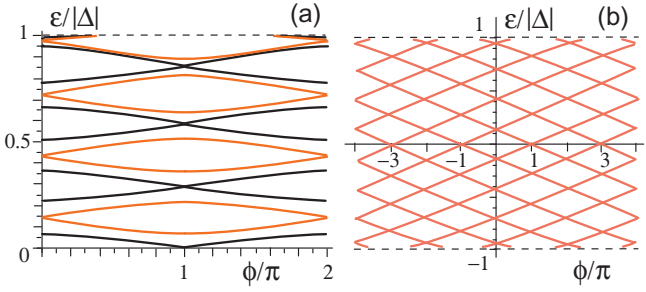}} %
%\begin{figure}[t] %h=here,t=top,b=bottom,p=erilliselle kuvasivulle
%\centerline{\epsfxsize= 6cm\epsfbox{trajectories.eps}}
\caption{Energy spectra Eq.\ (\protect\ref{determinant2}) for a
SINIS contact with $Z=0.5$ and $\Delta d/\hbar v_x =10$. (a) Dark
lines: In the resonance $\sin\alpha^\prime =0$; light (red online)
lines: in the anti-resonance $|\sin \alpha^\prime |=1$. (b) Pathes
connecting states below $-|\Delta|$ to those above $+|\Delta|$ for
trajectories $m=n=0$ of Fig.\ \protect\ref{fig-phialpha}.}
\label{fig-sns-long1}
\end{figure}

The normal conductor of the length $d\gtrsim v_x/|\Delta|$ is
chosen to have a single conducting mode; here $v_x$ is the Fermi
velocity of the conducting mode and $|\Delta|$ is the
superconducting gap in the leads. The energy spectrum of sub-gap
Andreev states, $\epsilon_n (\phi)$, has a large number $N\sim
|\Delta| d/\hbar v_x$ of levels each being a function of the phase
difference $\phi$ between the superconducting leads; a typical
spectrum is shown in Fig.\ \ref{fig-sns-long1}(a).  For a finite
strength $Z$ of barriers between the leads and the normal
conductor, the levels are separated by the minigaps $\Delta
\epsilon_{2\pi}$ at $\phi = 2\pi k$ and by minigaps $\Delta
\epsilon_{\pi}$ at $\phi = \pi (1+2k)$, where $k$ is an integer
\cite{bagwell92}. Magnitudes of these minigaps depend on the
interference phase $\alpha^\prime =k_xd+\delta$ between the waves
incident on and reflected from the barriers ($k_x$ is the Fermi
wave vector of the mode and $\delta$ is the scattering phase). The
minigaps $\Delta \epsilon_{\pi}$ disappear in the resonance
determined by the condition $\sin \alpha^\prime =0$. In its turn,
$\Delta \epsilon_{2\pi}$ disappear in the anti-resonance where
$\sin \alpha^\prime =\pm 1$. The Fermi velocity and thus the
interference phase $\alpha^\prime$ can be tuned by the gate
voltage $V_g$ as shown in Fig. \ref{fig-quantdot}. Thus one can
close either the gaps at $\phi =2\pi k$ or the gaps at $\phi =\pi
(1+2k)$ sequentially adjusting a proper time dependence of $V_g$.
Choosing the bias voltage $V=(\hbar /2e)d\phi /dt$  such that
$\phi =0$ at the time when the gap $\Delta \epsilon_{2\pi}$ is
closed and $\phi=\pi$ at the time when $\Delta \epsilon_{\pi}$ is
closed, and so on, a resonance energy and charge pumping via
spectral flow of excitations from states below $-|\Delta|$ to
states above $+|\Delta|$ and back will take place   realizing an
``Archimedean screw'' \cite{AltshulGlazm} in energy space. The
particles moving upwards will reach continuum at $\epsilon
=+|\Delta|$ with the distribution corresponding to the equilibrium
at $\epsilon =-|\Delta|$ and vice versa. The subsequent relaxation
of ``wrong'' distributions in continuum is accompanied by large
dissipation leading to a dc current component under a dc bias
voltage. It is essential that the gaps disappeared sequentially:
vanishing of just one set of gaps either at $2\pi k$ or at
$\pi(1+2k)$ does not lead to the energy or charge pumping. A
possible way to realize this device is to bridge two
superconductors by a carbon nanotube \cite{nanotubes} subject to a
gate voltage. A similar technology was used in Ref.\
\cite{Kouwenhoven} to design semiconductor nanowires with tunable
supercurrent.

%%%%%%%%%%%%%%%%%%%%%%%%%%%%%%%%%%%%%%%%%%%%%%%%%%%%%%%%%%%%%%%%%%%%%
%%%%%%%%%%%%%%%%%%%%%%%%%%%%%%%%%%%%%%%%%%%%%%%%%%%%%%%%%%%%%%%%%%%%%%

%%%%%%%%%%%%%%%%%%%

\begin{figure}[t]
\centerline{\includegraphics[width=0.6\linewidth]{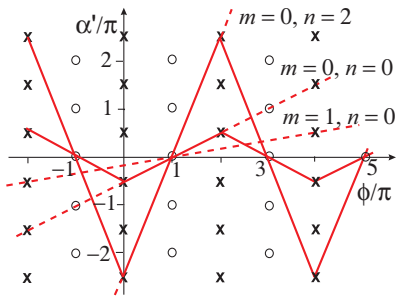}} %
%\begin{figure}[t] %h=here,t=top,b=bottom,p=erilliselle kuvasivulle
%\centerline{\epsfxsize= 6cm\epsfbox{trajectories.eps}}
\caption{ Linear resonance trajectories. } \label{fig-phialpha}
\end{figure}

{\it Resonances}.-- To close the mini-gaps in resonance with
$\phi$ one needs to tune the amplitude and frequency of the gate
voltage in accordance with the bias voltage. Let us plot in Fig.
\ref{fig-phialpha} two sets of points in the $\phi,\alpha ^\prime$
plane: (A) points $\phi = 2\pi k$, $\alpha ^\prime = \pi (\tilde k
+1/2)$ (shown by crosses) and (B) points $\phi = \pi (2 k +1)$,
$\alpha ^\prime = \pi \tilde k$ (circles). Resonance is achieved
when, changing $\alpha ^\prime$ and $\phi$ with time, one obtains
a trajectory that passes first through a point A, then through a
point B, then again through A, and so on. For linear trajectories
parametrized as $2\alpha ^\prime =\omega _gt+2\pi k$, $ \phi-\pi =
\omega_J t$, where $\omega_J = 2eV/\hbar$, this requires $
\omega_J/\omega_g =(1+2m)/(1+2n) $. Some of these trajectories are
shown in Fig.~\ref{fig-phialpha} by dashed lines. The resonance
trajectories open continuous paths in the $\epsilon, \phi$ space
connecting states below $-|\Delta|$ to those above $|\Delta|$, see
Fig.\ \ref{fig-sns-long1}(b). The gate voltage can be varied
within finite intervals along the trajectories [broken lines in
Fig.~\ref{fig-phialpha}] equivalent to the straight lines since
$\sin^2 \alpha^\prime =\sin^2 \omega_g t/2$. More practical is to
apply a sinusoidal gate voltage with a frequency $\Omega$. For
given $m$ and $n$, the trajectory passing through the same points
is
\begin{equation}
\alpha ^\prime \equiv (e d/\hbar v_x) V_g=[(1+2n)\pi /2]\sin
(\Omega t) \label{rescond-rational2}
\end{equation}
where $ \phi-\pi = \omega_J t$ with $\omega _J=2\Omega(1+2m)$; the
equivalent $\omega _g=2\Omega(1+2n)$. The amplitude of the gate
voltage should be odd rational of $\pi \hbar v_x/2ed $; note that
$e V_g \ll \Delta$ for long contacts with $N\gg 1$.

%%%%%%%%%%%%%%%%%%%%%%%%%%%

{\it Spectrum}.--We use the Bogoliubov--deGennes equations
\begin{equation}
\left[ -\frac{\hbar ^2}{2m}\frac{d^2}{dx^2}-E_{F} +U(x)\right]\hat
\sigma _z \hat\psi +\hat H \hat \psi  =\epsilon \hat \psi \ ,
\label{eqBog}
\end{equation}
to explicitly demonstrate the resonance properties of the sub-gap
energy spectrum of a SINIS structure. Here $\hat \sigma_z$ is the
Pauli matrix in Nambu space, and
\[
\hat \psi =\left(\begin{array}{c} u\\ v\end{array}\right)\, ,\;
\hat H=\left(\begin{array}{cc} 0&\Delta \\
\Delta ^* &0 \end{array}\right) \ .
\]
The superconducting gap is $\Delta =|\Delta|e^{\pm i\phi /2}$ for
$x>d/2$ and $x<-d/2$, respectively, while $\Delta =0$ for
$-d/2<x<d/2$. For simplicity we model the normal reflections at
the interfaces as being produced by $\delta$-function barriers $
U(x)=I\delta (x-d/2)+I\delta (x+d/2)$.

In the normal region the particle, $e^{\pm iq_+ x}$, and hole,
$e^{\mp iq_- x}$, waves have amplitudes $u^\pm$ and $v^\mp$,
respectively. The upper or lower signs refer to the waves
propagating to the right $\hat \psi ^>=\left(u^+, v^-\right)$ or
to the left $\hat \psi ^< =\left(u^-, v^+\right)$. The particle
(hole) momentum is $q_\pm =k_x\pm \epsilon/\hbar v_x$. Scattering
at the right and left barrier couples the amplitudes of incident
and reflected waves \cite{Beenakker}:
\begin{equation}
\hat \psi ^< _R=\hat S^R\hat \psi ^>_R\, , \; \hat \psi ^> _L=\hat
S^L\hat \psi ^<_L \, ; \; \hat S=\left( \begin{array}{cc} S_{11} & S_{12} \\
S_{21} & S_{22}\end{array}\right)\ . \label{S-R,L}
\end{equation}
The scattering matrices for the right and left barriers are $\hat
S^R=\hat S(\chi _2)$ and $\hat S^L=\hat S(\chi _1)$, respectively,
where $\chi _1=-\phi /2$ while $\chi _2=\phi /2$. Components of
the $\hat S$ matrix for $\delta$-like barriers and energies
$|\epsilon|<|\Delta|$ are \cite{BKT}
\begin{eqnarray*}
S_{11}(\chi,Z)=S_{22}(\chi,-Z)=-\frac{(U^2-V^2)(Z^2+iZ)}{U^2+(U^2-V^2)Z^2} \ ,
\\
S_{12}(\chi,Z)e^{-i\chi }=S_{21}(\chi,Z)e^{i\chi }
=\frac{UV}{U^2+(U^2-V^2)Z^2} \ .
\end{eqnarray*}
Here $Z=mI/\hbar ^2 k_x$ is the barrier strength and $
U=2^{-1/2}[1+i\sqrt{|\Delta|^2-\epsilon ^2}/\epsilon ]^{1/2}$,
$V=U^*$. The waves at different ends of the normal channel have
different phases
\begin{equation}
\hat \psi ^>_R=e^{i(\alpha \hat \sigma _z+\beta)}\hat \psi ^>_L \,
, \; \hat \psi ^<_R=e^{-i(\alpha \hat \sigma _z+\beta)}\hat \psi
^<_L \label{R-Lphases}
\end{equation}
where $ \alpha =k_xd\, , \; \beta = \epsilon d/\hbar v_x $. Using
unitarity of $\hat S$ matrix, $\hat S^\dagger \hat S =1$, the
condition of solvability of Eqs. (\ref{S-R,L}) and
(\ref{R-Lphases}) assumes a compact form
\begin{equation}
|S_{11}|^2\sin ^2 \alpha ^\prime +|S_{12}|^2\cos ^2(\phi /2)
=\sin^2(\beta +\gamma) \ . \label{determinant2}
\end{equation}
Here $\alpha ^\prime =\alpha +\delta $ and the scattering phase
$\delta$ is introduced through $ \cot \delta =Z $; the phase
$\gamma$ is defined as $ e^{2i\gamma}=S_{11}/S^*_{22}$. Equation
(\ref{determinant2}) determines the energy spectrum of a SINIS
contact. Such spectrum has been extensively studied by many
authors (see e.g. \cite{spectrum}).

%%%%%%%%%%%%%%%%%%%%%%%%%%%%%%%%%%%%%%%%%%%%%%%%%%%%%%%%%%%%%%%%%%%%%
%%%%%%%%%%%%%%%%%%%%%%%%%%%%%%%%%%%%%%%%%%%%%%%%%%%%%%%%%%%%%%%%%%%%%%

In what follows we focus on long contacts, $d\gg \hbar v_x/|\Delta
|$, with a large number of levels $N$. Examples of the spectra are
shown in Fig.~\ref{fig-sns-long1}(a). All gaps at $\phi =\pi (1+2
k)$ disappear at the same resonance parameter $\sin \alpha ^\prime
=0$. Similarly, all gaps at $\phi = 2\pi k$ disappear when $|\sin
\alpha ^\prime |=1$. This follows from Eq.\ (\ref{determinant2})
due to unitarity $|S_{11}|^2+|S_{21}|^2=1$. The low-energy levels,
$\epsilon_n \ll \Delta$, have the form $\epsilon_n =\pm \epsilon
_0 +\pi \hbar v_x n/d$ where
\begin{equation}
\epsilon_0 =\frac{\hbar v_x}{d} \arcsin \sqrt{{\cal T}^2\cos
^2\frac{\phi}{2}+(1-{\cal T}^2)\sin ^2\alpha ^\prime} \ ,
\label{spectrum-sym}
\end{equation}
$n$ is an integer, $ {\cal T}=(1+2Z^2)^{-1} $, and we use $\gamma
\ll 1$. The spectrum has energy gaps at $\phi =\pi (1+2 k)$
\[
\Delta \epsilon_{\pi}=(2\hbar v_x/d)\arcsin \left[ |\sin \alpha
^\prime |\sqrt{1-{\cal T}^2}\right] \ .
\]
The gaps $\Delta \epsilon_{2\pi}$ at $\phi =2\pi k$ are given by
the same expression with $|\sin \alpha ^\prime |$ replaced with
$|\cos \alpha ^\prime |$. For a transparent contact, ${\cal T}=1$,
minigaps disappear.

%%%%%%%%%%%%%%%%%%%%%%

{\it Current}.-- For time-dependent $E_g$ and $\phi$ the current
can be calculated \cite{Lundin} by varying Eqs.\ (\ref{eqBog})
with respect to $\delta \Delta$ and the chemical potential of the
normal conductor controlled by the gate voltage:
\[
\delta E_F \equiv \delta E_g(x)=\left\{ \begin{array}{cr} -e\delta
V_g, & |x|<d/2 \\ 0, & |x|>d/2\end{array} \right. \ .
\]
Note that variations of $\beta$, $\gamma$, and of the scattering
phase $\delta$ resulting from changes in $k_x$ are negligible. If
the magnitude $|\Delta|$ is constant, $ \delta \Delta =i\delta
\chi \Delta $, we find
\begin{equation}
(\hbar/2e)\left(\delta \chi _R J_{nR} -\delta \chi_L
J_{nL}\right)=\delta \epsilon _n -Q_n\delta V_g \ , \label{vareq4}
\end{equation}
where $ Q_n=e\int (|u_n|^2 -|v_n|^2) \, dx $ is the charge on
level $n$; the integral in $Q_n$ is extended only over the length
of the normal conductor since $ |u_n|^2-|v_n|^2=0$ in an open
superconducting region. We denote
\[
J_n=-(ie\hbar/2m)\left[ u^*_n (du_n/dx)  + v^*_n (dv_n/dx) - c.c.
\right]\ ,
\]
$J_{nR,L}$ are the values of $J_n$ near the right (left) barrier.

Let $\epsilon _n$, $\chi $, and $E_g$ adiabatically depend on time
as a parameter. Summing up the Andreev states having the
distribution function $f_n$ we find the energy balance
\cite{Lundin}
\begin{equation}
VI^{\rm sg} = \sum _n (1-2f_n)
 Q_n \frac{d V_g}{d t} - \sum _n
\frac{d\epsilon_n }{d t}(1-2f_n) \ . \label{curent-sum}
\end{equation}
$ I^{\rm sg}=\left(I_R^{\rm sg}+I_L^{\rm sg}\right)/2 $ is the
current carried by the sub-gap states, $ (\hbar/2) d \chi_{R,L}/d
t =-eV _{R,L}=\pm eV/2$ due to symmetry; $V_{R,L}$ and
$I_{R,L}^{\rm sg}$ are the potentials and currents in the leads.
In Eq.~(\ref{curent-sum}), the work done by the bias voltage
$VI^{\rm sg}$ and by the gate voltage (the first term in the
r.h.s.) produce the energy change of the Andreev states.

Keeping the phases fixed in Eq. (\ref{vareq4}) we find $Q_n =
\partial \epsilon _n/\partial V_g $. Since $\epsilon _n$
is a function of two parameters, $V_g$ and $\phi$, Eq.
(\ref{curent-sum}) contains only the partial derivative
\begin{equation}
I^{\rm sg}= - \frac{2e}{\hbar} \sum _n \frac{\partial \epsilon_n
}{\partial \phi}(1-2f_n) \label{current-final}
\end{equation}
where $ d \phi /d t=2eV/\hbar $ was used. This equation formally
coincides with the known time-independent result
\cite{Beenakker,Kupriyanov}.

Note that the average currents satisfy $\overline{I}_g\equiv
\overline{I}_R-\overline{I}_L=0$ where $I_g$ is the gate current
(see Fig.~\ref{fig-quantdot}). Therefore $\overline{I}$ is the
average current through each lead. It is determined by the longer
time of the particle drift through the levels while the average
gate current has also contributions due to much more rapid jumps
of the distribution function each time when one of the levels
joins the continuum, thus $\overline{I}_g \ne -\sum _n \dot
Q_n(1-2f_n)$.

%%%%%%%%%%%%%%%%%%%%%%%

{\it Charge pumping}.-- Consider the simplest case where the gate
voltage varies along the broken line in Fig.~\ref{fig-phialpha}
equivalent to a straight line $m=n=0$. Let the deviation from the
resonances be
\begin{equation}
\phi -\pi =\omega_J t\, , \; 2\alpha ^\prime =\omega_J t +2\alpha
_0 \ , \label{rescond1}
\end{equation}
where $\alpha_0$ is determined by initial conditions, and
calculate the average current as a function of $\alpha_0$. As the
phase point moves along the trajectory with $\alpha_0\ne 0$, the
distance between nearest levels does not become less than $\Delta
\epsilon $. For low energies from Eq.\ (\ref{spectrum-sym}) and
small $\alpha_0$,
\[
\Delta \epsilon =(\hbar v_x/d)\sqrt{4{\cal T}^2(1-{\cal
T}^2)\alpha _0^2+\omega_J ^2(t-t_0)^2} \ .
\]
The probability of Zener tunnelling between the levels is
\[
p=\exp \left[ -2\pi v_x{\cal T}^2(1-{\cal T}^2)\alpha
_0^2/\omega_J d\right]
\]
for all $\phi \approx \pi k$. The Andreev states Eq.\
(\ref{determinant2}) are well-defined under an adiabatic condition
that $p$ is small far from the avoided crossings, $ \omega_J \ll
v_x{\cal T}^2(1-{\cal T}^2)/d $.

Consider first an ideal adiabatic process when transitions between
the levels are absent and the distribution $f_n$ on each level is
constant. Equation (\ref{determinant2}) yields
\begin{equation}
\sin[2(\beta +\gamma)]\frac{\partial \beta }{\partial
\epsilon}\frac{\partial \epsilon}{\partial \phi}
=\frac{1}{2}|S_{12}|^2\sin(\phi -\pi) \ . \label{derivative-long}
\end{equation}
Note that $\partial \beta /\partial \epsilon = d/\hbar v_x $. The
derivatives $\partial \gamma /\partial \epsilon \sim
\partial |S|^2/\partial \epsilon \sim (\Delta^2-\epsilon^2)^{-1/2}
$ if $Z\sim 1$; they can be neglected for long contacts if
$\epsilon$ is not very close to $\Delta$. We put $\phi
=\phi^\prime +\pi$ in the average $\overline{(\partial \epsilon
_n/\partial \phi )}=(2\pi)^{-1}\int _0^{2\pi}(\partial \epsilon
_n/\partial \phi )\, d\phi$ and see that $\sin [2(\beta +\gamma)]$
in Eq.\ (\ref{derivative-long}) does not change its sign, while
$\partial \epsilon/\partial \phi$ does if the particle stays on
the same level. The average current vanishes for any $\alpha _0\ne
0$: Pumping is absent without interlevel transitions followed by
relaxation of the electronic distribution in continuum.

If $p$ is finite, the particles move up or down the spectrum while
the distribution relaxes in continuum due to fast escape through
the potential barriers. The latter requires $ \omega_J \ll
v_x{\cal T}/d $ which is not as strict as the adiabatic condition
above. The average current is now finite for nonzero $\alpha_0$.
When $p$ reaches $p=1$ for $\alpha_0=0$, a free drift through the
levels is realized along the continuous paths of Fig.\
\ref{fig-sns-long1}(b) in the $\epsilon, \phi$ space. The
distribution function at each level is again constant though its
deviation from equilibrium is the largest: For particles moving
down (up) the levels it coincides with the equilibrium
$1-2f_n=\pm\tanh (|\Delta |/2T)$ at $\epsilon =\pm |\Delta|$. At
this point, the dc current reaches its maximum. For a continuous
path Eqs.\ (\ref{determinant2}), (\ref{derivative-long}) yield  $
\beta +\gamma =\pm (\phi - \pi)/2 +\pi n$ and
\begin{equation}
\partial \epsilon/\partial \phi =\pm (\hbar v_x/2d)|S_{12}|^2 \ .
\label{deriv-res2}
\end{equation}
Transitions from one level to another at the crossing points $\phi
=\pi k$ are accompanied by sign changes of $\sin[2(\beta
+\gamma)]$ in Eq.\ (\ref{derivative-long}) while the sign of
$\partial \epsilon _n/\partial \phi$ is unchanged; in Eq.\
(\ref{deriv-res2}) it is plus for transitions upwards and minus
otherwise. Since the variation in energy for each level is small
compared to $\Delta$ we finally obtain from Eq.\
(\ref{current-final}) replacing the sum with the integral
\begin{eqnarray}
\overline{
I}_{max}&=&\frac{2ev_x}{d}\tanh\left(\frac{|\Delta|}{2T}\right)
\int_{-|\Delta|}^{|\Delta|}|S_{12}(\epsilon)|^2
\, \frac{dn}{d\epsilon}\, d\epsilon \nonumber \\
&=&\frac{8e|\Delta |}{\pi\hbar}\frac{{\cal T}^2L}{\sqrt{1-{\cal
T}^2}} \tanh\left(\frac{|\Delta|}{2T}\right) \ , \quad
\label{averagecurrent}
\end{eqnarray}
where $ L=\ln\left[\cot(\delta /2)\right] $. The density of states
is $ dn/d \epsilon=\pi ^{-1} d\beta /d\epsilon=d/\pi\hbar v_x $.

Equation (\ref{averagecurrent}) accounts for the dc current from
sub-gap levels (the corresponding superscript is omitted).
Continuum states do also contribute. Indeed, for low transparency
the Josephson current is \cite{Furusaki,GalZaik02} $ I_{J}=
[ev_x{\cal T}^2Y(\alpha ^\prime)/ \pi d]\sin \phi $ where
$Y(\alpha ^\prime)= (2\alpha ^\prime -\pi)/\sin (2\alpha^\prime
-\pi) $ and $0 <\alpha ^\prime <\pi $. Dc contributions of the
type of Shapiro steps appear when the transparency ${\cal
T}^2Y(\alpha ^\prime)$ is ac modulated in resonance with
$\omega_J$. However, the magnitude of these steps is proportional
to $ev_x{\cal T}^2/ d$ and is thus much smaller than the dc
current in Eq.~(\ref{averagecurrent}).

%%%%%%%%%%%%

{\it Discussion}.--It is instructive to look at the energy balance
Eq.\ (\ref{curent-sum}). Since $\phi -\pi =\omega_Jt$, the level
sum of separate averages of the second term in the r.h.s.\ can be
written as an integral over one continuous path $\epsilon (\phi)$
(see, Fig.\ \ref{fig-sns-long1}(b)).
 For particles moving upwards, $ \sum _n
\overline{d \epsilon_n /dt} =(2\pi)^{-1}\int_{\phi_-}^{\phi_+}
(d\epsilon /d \phi)\, d\phi $, where $\epsilon (\phi_{\pm})=\pm
|\Delta|$. Similarly to adiabatic pumps
\cite{AltshulGlazm,Buttiker} this integral can be viewed as a
circulation of a ``vector potential'' $d\epsilon_n /d \phi$ along
a closed contour in the plane of complex $\Delta$ (circle of
constant $|\Delta|$). Each continuous path provides a nonzero
circulation for an adiabatic process that takes place between two
dissipative events when the path merges with continuum. This term
thus gives a contribution similar to that of multiple Andereev
reflection processes in ballistic SNS junctions
\cite{GunsenZaik94}; in our case the barrier-induced minigaps are
closed by the resonance gate voltage. The work by the gate voltage
(first term) in Eq.~(\ref{curent-sum}) is an energy counterflow;
it reduces the pumped charge as compared to its ballistic value.
Indeed, $\sum_n\overline{Q_n (dV_g/dt)}= -(\pi \hbar
v_x/ed)\sum_n\overline{Q}_n$. The average current is thus
proportional to the Andreev probability, $-(\pi \hbar
v_x/ed)\sum_n (e+ \overline{Q}_n)=-(\pi \hbar v_x/d)\sum_n
|S_{12}|^2$ which agrees with Eq. (\ref{averagecurrent}).

The energy varies from $-|\Delta|$ to $+|\Delta|$ during time $
T_0 \sim |\Delta| d/\hbar v_F\omega_J$ which should be shorter
than the inelastic relaxation time $\tau _\epsilon$. Combined with
the adiabatic condition this gives the Josephson frequency window for the
first resonance, $ |\Delta| d/\hbar v_F \tau _\epsilon \ll
2eV/\hbar \ll v_F {\cal T}^2(1-{\cal T}^2)/d $. Such slow
$\tau_\epsilon^{-1}$ requires the use of low temperatures.

In the case of a sinusoidal gate voltage
Eq.~(\ref{rescond-rational2}), resonances also occur for rational
gate voltage amplitudes specified by $n$ and for $\omega_J
=2\Omega(1+2m)$. The dc current-voltage curve has peaks $\bar
I_m\sim \bar I_{max}/(1+2m)$ decaying with $m$ but weakly
depending on $n$. The frequency $2eV/\hbar(1+2m)$ should also
belong to the window above. This puts an upper limit on
observation of peaks of higher order in $m$. Ideally, the largest
peak $\overline{I}_{max}$ is $N$ times higher than the usual
Shapiro step.

To summarize, we found a new mechanism of nonequilibrium charge
pumping. It is based on the spectral flow in a voltage-biased
SINIS quantum junction when the minigaps in the energy spectrum
are closed sequentially in resonance with the Josephson frequency.

We thank Yu.\ Galperin and V.\ Kozub for valuable discussions.
This work was supported, in part, by the US DOE Office of Science,
contract No.\ W-31-109-ENG-38, by Russian Foundation for Basic
Research, by Russian Science Support Foundation, and by Program
``Quantum Macrophysics'' of RAS. ASM acknowledges the support by
the Academy of Finland.


\begin{thebibliography}{99}

\bibitem{AltshulGlazm} B.L. Altshuler, L.I. Glazman, Science {\bf
283}, 1864 (1999).
\bibitem{Switkes} M. Switkes, C.M. Marcus, K. Campman, A.C.
Gossard, Science {\bf 283}, 1905 (1999).

\bibitem{Buttiker} M. Moskalets and M. B{\"u}ttiker, Phys. Rev. B
{\bf 72}, 035324 (2005); M. B{\"u}ttiker, J.\ Low Temp.\ Phys.\
{\bf 118}, 519 (2000).

\bibitem{Hekking} M. Governale, F. Taddei, R. Fazio, and F.W.J. Hekking,
cond-mat/0506078.

\bibitem{bagwell92}
P.\ F.\ Bagwell, Phys.\ Rev.\ B\ {\bf 46}, 12573 (1992).

\bibitem{nanotubes} M.R.\ Buitelaar, W.\ Belzig, T.\ Nussbaumer,
et al.,  Phys.\ Rev.\ Lett.\ {\bf 91}, 057005 (2003); A.\ Kosumov,
M.\ Kociak, M.\ Ferrier, et al., Phys.\ Rev.\ B {\bf 68}, 214521
(2003).

\bibitem{Kouwenhoven} Y.J.\ Doh, J.A.\ van Dam , A.L.\ Roest, et
al., Science {\bf 309}, 272 (2005).

\bibitem{Beenakker} C.W.J.\ Beenakker, Phys.\ Rev.\ Lett.\ {\bf 67},
3836, (1991).

\bibitem{BKT} G.E.\ Blonder, M.\ Tinkham, and T.M.\ Klapwijk, Phys.\
Rev.\ B, {\bf 25}, 4515 (1982).

\bibitem{spectrum} U.\ Sch{\"u}ssler and R.\ K{\"u}mmel, Phys.\ Rev.\ B {\bf 47},
2754 (1993); G.A.\ Gogadze and A.M.\ Kosevich, Fiz.\ Nizkih Temp.\
{\bf 24}, 716 (1998) [Low Temp.\ Phys.\ {\bf 24}, 540 (1998)]; A.\
Jacobs and R.\ K{\"u}mmel,  Phys.\ Rev.\ B {\bf 71}, 184504
(2005); D.D.\ Kuhn, N.M.\ Chtchelkatchev, G.B.\ Lesovik, and G.\
Blatter, Phys.\ Rev.\ B, {\bf 63}, 054520 (2001).

\bibitem{Lundin} N.I.\ Lundin, L.Y.\ Gorelik, R.I.\ Shekhter, M.\
Jonson, Superlattices and Microstructures, {\bf 25}, 937 (1999).

\bibitem{Kupriyanov} A.A.\ Golubov, M.Yu.\ Kupriyanov and E.\
Il'ichev, Rev.\ Mod.\ Phys.\ {\bf 76}, 411 (2004).

\bibitem{Furusaki} A.\ Furusaki, H.\ Takayanagi, and M.\ Tsukada, Phys.\ Rev.\ B
{\bf 45}, 10563 (1992).

\bibitem{GalZaik02} A.V.\ Galaktionov and A.D.\ Zaikin, Phys.\ Rev.\ B
{\bf 65}, 184507 (2002).

\bibitem{GunsenZaik94} U.\ Gunsenheimer and A.D.\ Zaikin,
Phys.\ Rev.\ B {\bf 50}, 6317 (1994).


\end{thebibliography}
\end{document}